\documentclass[article,nojss]{jss}
\usepackage{orcidlink,thumbpdf,lmodern}
\graphicspath{{Figures/}}

\usepackage{amsmath,amsthm,amssymb,bm,booktabs}
\newtheorem{theorem}{Theorem}
\theoremstyle{plain}
\newtheorem{example}[theorem]{Example} 

\author{Bettina Gr\"un\\WU Vienna University\\of Economics and Business
  \And
  Gertraud Malsiner-Walli\\WU Vienna University\\of Economics and Business}

\title{Bayesian Finite Mixture Models}

\Plainauthor{Bettina Gr\"un, Gertraud Malsiner-Walli}

\Address{
  Bettina Gr\"un\\
  Institute for Statistics and Mathematics\\
  WU Vienna University of Economics and Business\\
  Welthandelsplatz 1, 1020 Wien, Austria\\
  E-mail: \email{bettina.gruen@wu.ac.at} \\
  
  Gertraud Malsiner-Walli \\
  Institute for Statistics and Mathematics\\
  WU Vienna University of Economics and Business\\
  Welthandelsplatz 1, 1020 Wien, Austria\\
  E-mail: \email{gertraud.malsiner-walli@wu.ac.at}
}

\Abstract{
  Finite mixture models are a useful statistical model class
  for clustering and density approximation.  In the Bayesian framework
  finite mixture models require the specification of suitable priors
  in addition to the data model. These priors allow to avoid spurious
  results and provide a principled way to define cluster shapes and a
  preference for specific cluster solutions. A generic model
  estimation scheme for finite mixtures with a fixed number of
  components is available using Markov chain Monte Carlo (MCMC)
  sampling with data augmentation. The posterior allows to assess
  uncertainty in a comprehensive way, but component-specific posterior
  inference requires resolving the label switching issue.

  In this paper we focus on the application of Bayesian finite mixture
  models for clustering.  We start with discussing suitable
  specification, estimation and inference of the model if the number
  of components is assumed to be known. We then continue to explain
  suitable strategies for fitting Bayesian finite mixture models when
  the number of components is not known.  In addition, all steps
  required to perform Bayesian finite mixture modeling are illustrated
  on a data example where a finite mixture model of multivariate
  Gaussian distributions is fitted. Suitable prior specification,
  estimation using MCMC and posterior inference are discussed for this
  example assuming the number of components to be known as well as
  unknown.
}


\Disclaimer{This is a preprint of an article published in
\emph{Wiley StatsRef: Statistics Reference Online}.  Please cite as:
Bettina Gr{\"u}n and Gertraud Malsiner-Walli (2022) Bayesian finite
mixture models. In N. Balakrishnan, Theodore Colton, Brian Everitt,
Walter Piegorsch, Fabrizio Ruggeri, and Jef L. Teugels, editors,
\emph{Wiley StatsRef: Statistics Reference Online}
\doi{10.1002/9781118445112.stat08373}.}

\begin{document}

\section{Introduction}\label{sec:introduction}

Finite mixture models have a long history in statistics with their
first use proposed by \citet{Newcomb:1886} and \citet{Pearson:1894}
more than hundred years ago. They represent a useful model class for
cluster analysis as well as semi-parametric density estimation. In
addition the model class enjoys great flexibility and extensibility
because arbitrary statistical models can be incorporated in a mixture
which corresponds to fitting a convex combination of these models to
the data. 

\citet{McLachlan+Peel:2000} and \citet{Fruehwirth-Schnatter:2006}
provide a thorough and detailed introduction into finite mixture
models discussing different extensions and their application.
\citet{McLachlan+Peel:2000} focus on maximum likelihood estimation,
while \citet{Fruehwirth-Schnatter:2006} discusses the Bayesian
estimation in detail. A more recent overview and discussion of new
developments in mixture analysis are compiled in
\citet{Fruehwirth-Schnatter:Celeux+Robert:2019}. This contribution
focuses on Bayesian estimation of finite mixture models in a
model-based clustering context.

The application of finite mixture models in standard data analysis has
been impeded due to the difficulties faced when estimating and
performing inference. Their uptake has been abetted by the increase in
available computing power. Routine estimation of different kinds of
mixture models has become feasible using the expectation-maximization
algorithm \citep{Dempster+Laird+Rubin:1977} for maximum likelihood
estimation and Markov chain Monte Carlo methods in combination with
data augmentation \citep{Diebolt+Robert:1994} for Bayesian
estimation. Both these estimation methods represent a general
estimation framework equally allowing for general extensibility and
inclusion of arbitrary statistical models in the mixture.  

Using the Bayesian framework to specify and estimate a finite mixture
model has several advantages:
\begin{itemize}
\item Domain knowledge can be included in a principled way using
  suitable prior specifications.  When using mixture models for
  cluster analysis, the priors allow to specify the prototypical shape
  of the clusters \citep{Hennig:2015}. As an explanatory analysis
  tool, cluster analysis aims at detecting interesting structures in
  the data. The priors may guide the estimation procedure to focus on
  mixture models with certain characteristics. E.g., the priors
  specified can reflect and support aims such as obtaining a cluster
  solution with equally sized clusters or clusters which differ
  primarily in their centroids.

\item The mixture likelihood is highly irregular due to
  identifiability issues and multimodality as well as the presence of
  spurious modes. The specification of priors in Bayesian estimation
  allows to obtain a regularized and smoothed version of the mixture
  likelihood as posterior. In this way problems in maximum likelihood
  estimation can be avoided where the application of optimization
  methods suffers from this irregularity.

\item Parameter uncertainty can be easily assessed using the whole
  posterior distribution.  No reliance on asymptotic normality is
  required, allowing for valid inference in cases where regularity
  conditions are violated, e.g., for small data sets and mixtures with
  small component weights.
\end{itemize}

Despite these advantages, Bayesian finite mixture models are not
routinely applied yet because of uncertainty regarding the
specification of suitable priors and lack of understanding of their
impact \citep{Aitkin:2001}. Also issues with estimation as well as
inference are reported due to computational challenges
\citep{Celeux+Kamary+Malsiner-Walli:2019}. In the following all steps
of a Bayesian finite mixture modeling application are explained in
detail, including data model and prior specification, estimation and
inference. This is done for both cases: if the number of components is
assumed to be known as well as if this number is unknown. In addition
to outlining the methodology and discussing theoretical
considerations, a real data set serves as an example to illustrate the
application of each of the steps and as a basis to facilitate the
uptake of Bayesian estimation of finite mixture models in applied
research.

\section{Known Number of Components}\label{sec:known-numb-comp}

\subsection{Model Specification}\label{sec:model-specification}

\subsubsection{Data Model}

When using finite mixture models for clustering the usual assumption
is that a uni- or multivariate continuous or discrete outcome variable
$\bm{y}$ is sampled from a heterogeneous population where $K$ groups
are present.  Each group follows its own group-specific parametric
distribution.  Each observation in the sample belongs to one of the
$K$ groups, but the group memberships are not observed. It is further
assumed that the group sizes in the sample correspond to the
prevalence of the groups in the population.

Marginally, the finite mixture density of an observation $\bm{y}_i$ is
given by a convex combination of the group-specific parametric
densities with the weights corresponding to the group sizes:
\begin{align*}
  h(\bm{y}_i | \bm{\vartheta}_K) &= \sum_{k=1}^K \eta_k f(\bm{y}_i | \bm{\theta}_k),
\end{align*}
with the group sizes also referred to as component weights
\begin{align*}
  \eta_k & > 0, \, k=1,\ldots, K,& \sum_{k=1}^K \eta_k = 1
\end{align*}
and the group-specific or component density $f(\cdot | \bm{\theta})$
being a parametric density for $\bm{y}$ with component-specific
parameters $\bm{\theta}$. $\bm{\vartheta}_K$ denotes the complete set
of parameters of a finite mixture model with $K$ components consisting
of the component weights $\bm{\eta}_K = (\eta_1, \ldots, \eta_K)$ and
the component-specific parameters $\bm{\theta}_k$, $k=1,\ldots,K$
which are combined to
$\bm{\Theta}_K = (\bm{\theta}_k)_{k=1,\ldots,K}$.

Assuming that the observations $\bm{y}_i$, $i=1,\ldots,N$, are
independently identically distributed, the likelihood of the data
$\bm{y} = (\bm{y}_i)_{i=1,\ldots,N}$ is given by
\begin{align*}
  p(\bm{y}| \bm{\vartheta}_K) &= \prod_{i=1}^N \left[ \sum_{k=1}^K \eta_k f(\bm{y}_i | \bm{\theta}_k) \right]. 
\end{align*}
In contrast to other model likelihoods the mixture likelihood has the
property that the product over the observations is taken over sums
which implies for the log-likelihood that a sum of the log of sums is
determined. This complicates maximum likelihood estimation.

Other issues with the mixture likelihood are the multimodality of its
surface and the potential presence of spurious modes due to
unboundedness for parameters at the border of the parameter space.
Multimodality is caused by the invariance of the likelihood to
relabeling the components i.e., the likelihood value does not change
if the components are re-ordered. As a consequence, the mixture
likelihood function has $K!$ different, but equivalent modes
corresponding to all different ways of labeling.  Unbounded values
occur for example for multivariate mixtures of Gaussians in case the
mixture distribution contains a component where the mean value is
equal to an observed value $\bm{y}_i$ and the variance-covariance
matrix is the zero matrix.

The structure of the mixture likelihood is complicated because the
group memberships are not observed.  If each observation would not
only consist of $\bm{y}_i$ but also the group membership
$S_i \in \{1, \ldots, K\}$, the likelihood for these complete
observations $(\bm{y}_i, S_i)_{i=1,\ldots,N}$, which consist of the
observed data and the missing or unobserved data, would be simpler in
structure. The complete-data likelihood is given by
\begin{align*}
  p(\bm{y}, \bm{S} | \bm{\vartheta}_K) &= \prod_{i=1}^N \left[ \eta_{S_i} f(\bm{y}_i | \bm{\theta}_{S_i}) \right]. 
\end{align*}
The complete-data likelihood is easier to maximize.
\citet{Dempster+Laird+Rubin:1977} propose to use the
expectation-maximization algorithm for maximum likelihood or maximum
a-posteriori estimation exploiting the simplicity of the complete-data
log-likelihood compared to the observed-data log-likelihood.

\begin{example}[Data set and mixture model specification]\label{ex:data-model}
  The \code{diabetes} data set \citep{Reaven+Miller:1979} is used in
  \citet{Scrucca+Fop+Murphy:2016} to illustrate the use of finite
  mixture models in a maximum likelihood context. 
  The data set contains observations for 145 non-obese adults who were
  classified into three clinical groups (\code{class}) of diabetes ("Normal", "Overt", "Chemical"). The
  following measurements are available for the patients:
  \begin{itemize}
  \item \code{glucose}: the area under the plasma glucose curve after a three hour oral glucose tolerance test (OGTT),
  \item \code{insulin}: the area under the plasma insulin curve after a three hour OGTT,
  \item \code{sspg}: the steady state plasma glucose level.
  \end{itemize}
  \begin{figure}[t!]
  \centerline{\includegraphics[width=0.6\textwidth]{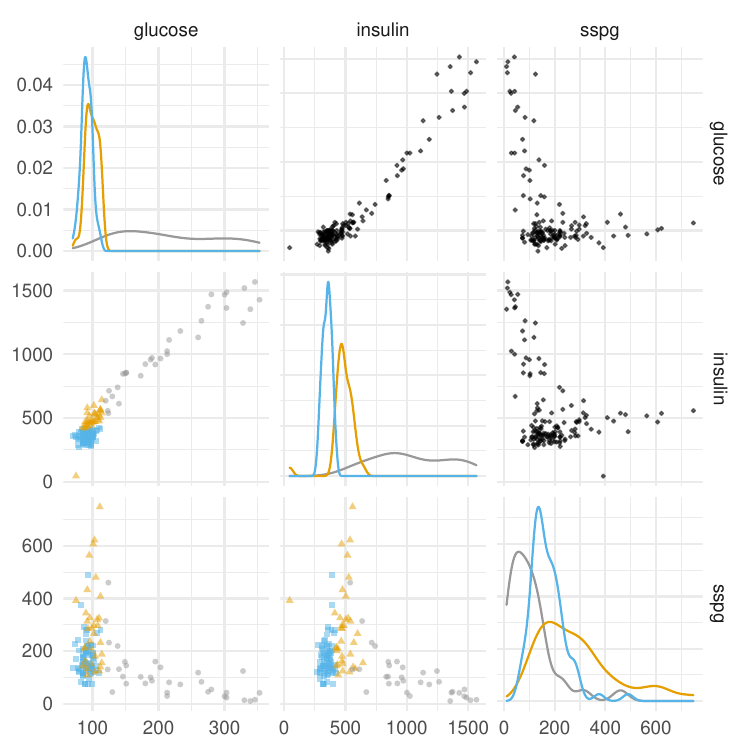}}
  \caption{Bivariate scatter plots of the clinical measurements in
    the data set \code{diabetes} in the off-diagonal with coloring according to the clinical
    classification in the lower triangle and kernel density estimates grouped by clinical
    classification in the diagonal.
    \label{fig:data}}
\end{figure}
The data is visualized in Figure~\ref{fig:data}. The upper triangle
contains scatter plots of the observed data $\bm{y}$ which consists of
three continuous variables \code{glucose}, \code{insulin} and
\code{sspg}.  The unobserved clinical group memberships are indicated
in the pairwise scatter plots in the lower triangle by color and
shape. The diagonal contains for each variable kernel density
estimates of the univariate distributions separately for each clinical
group.

In the following it is assumed that the data generating process for
this data set is a finite mixture distribution with three components,
each corresponding to one of the three clinical groups.  We assume
further that the observations within each group follow a multivariate
normal distribution, i.e., the component distribution is a
three-dimensional multivariate Gaussian distribution.  We will fit a
normal mixture model with three components to assess if the clinical
groups can be inferred when observing only the three clinical
measurements. The plot clearly indicates that, in order to capture the
groups, the component distributions need to differ in their mean as
well as their variance-covariance parameters and that using diagonal
variance-covariance matrices for the component distributions would not
provide a suitable fit. Thus in this section, where the number of
components $K$ are assumed to be known, the following data
generating process corresponding to a normal mixture model is imposed:
\begin{align*}
p(\bm{y} | \bm{\eta}_3,\bm{\mu}_1,\bm{\mu}_2,\bm{\mu}_3,\bm{\Sigma}_1,\bm{\Sigma}_2,\bm{\Sigma}_3) &= \sum_{k=1}^3 \eta_k f_N(\bm{y} | \bm{\mu}_k,\bm{\Sigma}_k),
\end{align*}
where $f_N(\cdot | \bm{\mu}, \bm{\Sigma})$ is the multivariate
Gaussian distribution density with mean $\bm{\mu}$ and
variance-covariance matrix $\bm{\Sigma}$.
\end{example}

\subsubsection{Prior Specification}

In the Bayesian framework, all information contained in the data
$\bm{y}$ about the parameters $\bm{\vartheta}_K$ is summarized in
terms of the posterior density $p(\bm{\vartheta}_K|\bm{y})$, which is
derived by combining the likelihood with the prior on the parameters
$p(\bm{\vartheta}_K)$ using Bayes’ theorem:
\begin{align*}
p(\bm{\vartheta}_K|\bm{y}) & \propto p(\bm{y}|\bm{\vartheta}_K)p(\bm{\vartheta}_K).
\end{align*}

The parameters $\bm{\vartheta}_K$ in a finite mixture model with $K$
components consist of the component weights
$\bm{\eta}_K = (\eta_1,\ldots,\eta_K)$ and the component-specific
parameters $\bm{\Theta}_K = (\bm{\theta}_k)_{k=1,\ldots,K}$.
%

The prior specification constitutes a crucial part of the finite
mixture model definition as only the joint specification of the data
model and the prior determines the statistical model.  No conjugate
prior exists for the mixture likelihood. This implies that no prior
specification is possible which would allow to obtain the posterior in
closed form. Thus, usually Markov chain Monte Carlo (MCMC) methods are
used to approximate the posterior.

Despite the need to use MCMC methods for posterior inference anyway,
 simplifying assumptions are in general made when specifying
priors for finite mixture models. In particular, in the following we
will only consider priors which have the following characteristics:
\begin{itemize}
\item The priors on the component weights and the component-specific
  parameters are assumed to be independent.  The component-specific
  parameters may also depend on some hyper-parameters $\bm{\phi}$ but
  are conditionally independent. This implies the following prior
  structure:
\begin{align*}
p(\bm{\eta}_K, \bm{\Theta}_K, \bm{\phi}) &=
p(\bm{\eta}_K) p(\bm{\Theta}_K | \bm{\phi}) p(\bm{\phi}) =
p(\bm{\eta}_K) \prod_{k=1}^K p(\bm{\theta}_k | \bm{\phi}) p(\bm{\phi}).
\end{align*}
Assuming independence simplifies the specification of the prior, in
particular given that usually no prior knowledge is available which
would suggest a specific dependency structure. Hyper-parameters on the
component-specific parameters allow to dilute the influence of the
specific priors imposed on the component-specific parameters and adapt
them to the data by combining information across the components. Again
this is a useful strategy in case no clear prior knowledge is
available which can be exploited to specify particular priors on the
component-specific parameters.
	
\item Proper priors are imposed on the component-specific parameters.
  Improper priors on the component-specific parameters lead to
  improper posteriors. This can easily be seen through the following
  considerations: The posterior is proportional to the likelihood
  times prior and the likelihood of $\bm{y}$ given $\bm{\vartheta}_K$
  can also be determined by summing the complete-data likelihood of
  $\bm{y}$ and $\bm{S}$ given $\bm{\vartheta}_K$ over all possible
  assignment vectors $\bm{S}$. In this sum, assignment vectors
  $\bm{S}$ which do not assign observations to all components lead to
  empty components where the posterior is identical to the
  prior. Hence an improper prior leads to an improper posterior.
\item The priors are label-invariant.  This implies that a-priori the
  distribution of each component is the same. Having label-invariant
  priors in combination with the label-invariant likelihood leads to a
  posterior which is also label-invariant. This is a sensible approach
  in case no clear prior information is available which would allow to
  differently characterize the components of the mixture.
  
\item The priors are conditionally conjugate.  This implies that,
  conditional on the data, the latent variables and all other
  parameters, the marginal posteriors of the parameters are available
  in closed form. This facilitates the application of a Gibbs sampling
  scheme in MCMC sampling for inference. In addition to the
  computational advantages this also leads to the use of well known
  parametric priors where the characteristics are well understood.
  While this specification restricts the parametric families
  considered for the priors, these priors still offer great
  flexibility by varying the parameter values or adding a
  hyper-prior. For an empirical example illustrating the influence of
  different parameter values on the results obtained see
  \cite{Gruen+Malsiner-Walli+Fruehwirth-Schnatter:2021}.
\end{itemize}

Using priors with these characteristics implies that the prior for the
component weights is specified independently of the component-specific
distribution and hence also is the same regardless of the specific mixture
model. To select the prior on the component weights $\bm{\eta}_K$, one
uses that the unobserved group memberships or assignment vectors
$\bm{S}$ are drawn from a multinomial distribution. The conjugate
prior is the Dirichlet distribution. To obtain a label-invariant prior
a symmetric $K$-dimensional Dirichlet distribution is specified which
depends only on a single scalar parameter $\gamma_K$:
\begin{align*}
\bm{\eta}_K &\sim \mathcal{D}(\gamma_K).
\end{align*}
The prior mean of the component weights is thus label-invariant and
given by $1/K$. The parameter $\gamma_K$ influences the prior
variability of the component weights with a prior variance of
$(K - 1)/(K^2 (K \gamma_K + 1))$.

The priors for the component-specific parameters $\bm{\theta}_k$,
$k=1,\ldots,K$ depend on the component-specific distribution of the
mixture.  They can be selected as the usual conjugate priors
of the component-specific distribution ignoring the mixture
structure. For the prior on the hyper-parameters $\bm{\phi}$ the
conjugate prior for the set of component-specific parameters is an
obvious choice.

\begin{example}[Prior specification]\label{ex:hier-gener-model}
  To fit a finite mixture of multivariate Gaussian distributions with
  three components to the \code{diabetes} data set, we need to specify
  $\gamma_K$ and priors for the component-specific means and the
  component-specific variance-covariance matrices.
  We choose $\gamma_K = 1$ which corresponds to imposing a uniform
  prior on the weights. If the number of components is known, this
  parameter is usually not influential.
	
  For the component-specific parameters we will use conditionally
  conjugate priors, i.e., a normal prior on the component means and an
  inverse Wishart prior on the variance-covariance matrices with a
  hierarchical prior on the second parameter. This specification is
  similar to the specification proposed in
  \citet{Richardson+Green:1997} for univariate Gaussian mixtures. This
  implies:
  \begin{align*}
    \bm{\mu}_k &\sim \mathcal{N}(\bm{b}_0, \bm{B}_0), k=1, \ldots, K,\\
    \bm{\Sigma}_k &\sim \mathcal{W}^{-1}(c_0, \bm{C}_0),  k=1, \ldots, K,\\
    \bm{C}_0 &\sim \mathcal{W}(g_0, \bm{G}_0).
  \end{align*}
  Note that we use the parameterization of the Wishart and inverse
  Wishart distribution as employed by
  \citet{Fruehwirth-Schnatter:2006}. This means that the density of
  $\bm{Y} \sim \mathcal{W}(\alpha, \bm{V})$ is given by
  \begin{align*}
    f(\bm{Y}| \alpha, \bm{V})&=\frac{|\bm{V}|^\alpha}{\Gamma(\alpha)}|\bm{Y}|^{\alpha-(r+1)/2}\exp\{-\text{tr}(\bm{V}\bm{Y})\},
  \end{align*}
  and of $\bm{Y} \sim \mathcal{W}^{-1}(\alpha, \bm{V})$ by
  \begin{align*}
    f(\bm{Y}| \alpha, \bm{V})&=
  \frac{|\bm{V}|^\alpha}{\Gamma(\alpha)}|\bm{Y}^{-1}|^{\alpha+(r+1)/2}\exp\{-\text{tr}(\bm{V}\bm{Y}^{-1})\},
  \end{align*}
  where
  $\Gamma(\alpha) = \pi^{r(r-1)/4}\prod_{j=1}^r \Gamma(\frac{2\alpha +
    1 -j}{2})$ and $r$ is the dimension of the observations, i.e.,
  $\bm{y}_i \in \mathbb{R}^r$.
  
  We select the prior parameters $\bm{b}_0$ and $\bm{B}_0$ to ensure
  that the analysis is invariant to the scaling of the data and to
  have proper priors which are barely informative to induce no
  shrinkage on the component-specific means to a common mean value. In
  particular we set $\bm{b}_0=\text{median}(\bm{y})$, i.e., equal to
  the component-wise median, and $\bm{B}_0=\bm{R}^2$, i.e., equal to
  the squared data range, in order to be very uninformative in regard
  to $\bm{b}_0$.
	
  Further, we define the degree of freedom parameters of the Wishart
  and inverse Wishart distribution to depend on the dimension $r$ of
  the data $\bm{y}_i \in \mathbb{R}^r$:
  \begin{align*}
    c_0 &= c + \frac{r+1}{2},&
    g_0 &= 1 + \frac{r-1}{2},
  \end{align*}
  where $c$ is a constant.  $\bm{C}_0$ is modeled in order to
  incorporate knowledge about the a-priori expected component
  variance-covariance matrix $\bm{\Sigma}_k$. To be scale-invariant
  the prior mean of $\bm{\Sigma}_k$ is assumed to be a fraction $\phi$
  of the diagonal matrix $\bm{S}$, which contains the
  diagonal values of the empirical variance-covariance matrix of the
  data $\textsf{COV}(\bm{y})$ in the diagonal.  Specifically, for the following
  specifications
  \begin{align*}
    \bm{C}_0 &= c\phi \bm{S},&
    \bm{G}_0 &= g_0 \bm{C}_0^{-1},
  \end{align*}
  the prior mean value of $\bm{\Sigma}_k$ is $\phi\bm{S}$.  A sensible
  choice for $\phi$ is, e.g., $\phi=0.75$.  The values of $c$ and
  $\phi\bm{S}$ can be interpreted as adding $c$ virtual observations
  with variance-covariance matrix $\phi\bm{S}$ to each group $k$.  If
  $c=2.5$, the eigenvalues of $\bm{\Sigma}_k\bm{\Sigma}_j^{-1}$ are
  bounded away from zero avoiding spurious modes in the posterior
  \citep[p.~192]{Fruehwirth-Schnatter:2006}.
\end{example}

\subsubsection{Hierarchical Generative Model}
Combining the data model with the prior specifications, the
hierarchical generative model for the parameters $\bm{\vartheta}_K$,
the unobserved assignment vectors $\bm{S}$ and the observed data
$\bm{y}$ is given by:
\begin{align} \label{eq:BFM}
\bm{\eta}_K |K,\gamma_K &\sim \mathcal{D}(\gamma_K),\\
\bm{\phi} &\sim p(\bm{\phi}),\\
\bm{\theta}_{k}|\bm{\phi} &\sim p(\bm{\theta} _{k}|\bm{\phi}),   \text{ independently for } k=1,\ldots,K, \\
S_i|K ,\bm{\eta}_K &\sim \mathcal{M}(1;\bm{\eta}_K),  \text{ independently for } i=1, \ldots,N ,\\
  \bm{y}_i|K,S_i=k,\bm{\theta}_k &\sim f(\bm{y}_i|\bm{\theta}_{k}), \text{ independently for } i=1, \ldots,N,
\end{align}
where $\mathcal{D}(\gamma_K)$ is the symmetric $K$-dimensional
Dirichlet distribution with scalar parameter $\gamma_K$ and
$\mathcal{M}(1;\bm{\eta}_K)$ is the multinomial distribution with one
trial and vector of success probabilities $\bm{\eta}_K$ for the $K$
possible outcomes.

\subsection{Model Estimation}\label{sec:model-estimation}

Different computational solutions for estimating Bayesian finite
mixture models with a fixed number of components $K$ and obtaining an
approximation of the posterior of $\bm{\vartheta}_K$ are considered in
\citet{Celeux+Kamary+Malsiner-Walli:2019}. A generic approach builds
on data augmentation and includes the component assignment vectors
$\bm{S}$ in the sampling scheme \citep{Diebolt+Robert:1994}. Using
conditionally conjugate priors, the resulting MCMC sampling scheme
only requires Gibbs steps. More details on MCMC methods and the
  Metropolis--Hastings algorithm in general are given in
  [\textbf{stat08285}. \textbf{stat07834}].

Given some initial parameter values $\bm{\eta}_K$ and $\bm{\Theta}_K$,
iterate Steps~1 and 2:
\begin{enumerate}
\item Classification of each observation $\bm{y}_i$, $i = 1, \ldots,N$
  conditional on $\bm{\eta}_K$ and $\bm{\Theta}_K$:
  \begin{enumerate}
  \item For $i=1,\ldots,N$: sample $S_i$ from
    \begin{align*}
      P(S_i=k|\bm{y}_i,\bm{\vartheta}_K) \propto \eta_k
      f(\bm{y}_i|\bm{\theta}_k).
    \end{align*}
  \end{enumerate}
\item Parameter simulation conditional on the classification $\bm{S}=(S_1, \ldots ,S_N)$:
  \begin{enumerate}
  \item Sample the mixture weights $\bm{\eta}_K$ from
    \begin{align*}
      \bm{\eta}_K | \bm{S} \sim \mathcal{D}(e_1,\ldots,e_K),
    \end{align*}
    where
    \begin{align*}
      e_k=\gamma_K+N_k 
    \end{align*}
    with $N_k=\#\{i:S_i=k\}$ the number of observations assigned to
    component $k$.
  \item Sample the component-specific parameters $\bm{\theta}_k$,
    $k=1,\ldots,K$ conditional on $\bm{y}$, $\bm{S}$ and $\bm{\phi}$ from their conditional conjugate posteriors.
  \item Sample hyper-parameters $\bm{\phi}$ conditional on
    $\bm{\Theta}_K$.
  \end{enumerate}
\end{enumerate}

\begin{example}[MCMC sampling scheme]\label{ex:model-estimation}
  To fit the finite mixture of multivariate Gaussian distributions to
  the \code{diabetes} data set, Steps~2(b) and 2(c) are specifically
  given by:
  \begin{enumerate}
  \item[2(b)] For all $k = 1,\ldots, K$, sample
    \begin{align*}
      \bm{\mu}_k | \bm{y}, \bm{S}, \bm{\Sigma}_k&\sim \mathcal{N}(\bm{b}_k, \bm{B}_k),\\
      \bm{\Sigma}_k^{-1} | \bm{y}, \bm{S}, \bm{\mu}_k &\sim \mathcal{W}(c_k, \bm{C}_k),
    \end{align*}
    with 
    \begin{align*}
      \bm{b}_k &= \bm{B}_k(\bm{B}_0^{-1}\bm{b}_0 + \bm{\Sigma}_k^{-1} N_k \bar{\bm{y}}_k),&
      \bm{B}_k &= (\bm{B}_0^{-1} +  N_k\bm{\Sigma}_k^{-1})^{-1},\\
      c_k &= c_0 + \frac{N_k}{2},&
      \bm{C}_k &= \bm{C}_0 + \frac{1}{2}\sum_{i:S_i = k}(\bm{y}_i - \bm{\mu}_k)(\bm{y}_i - \bm{\mu}_k)^{\top},                                  
    \end{align*}
    where
    $\bar{\bm{y}}_k = \frac{1}{N_k} \sum_{i: S_i = k} \bm{y}_i$.

  \item[2(c)] Sample
    \begin{align*}
      \bm{C}_0 | (\bm{\Sigma}_k)_{k=1,\ldots,K} \sim \mathcal{W}(g_0 + Kc_0, \bm{G}_0 + \sum_{k=1}^K \bm{\Sigma}_k^{-1}).
    \end{align*}
  \end{enumerate}
  See \citet[][Chapter~6]{Fruehwirth-Schnatter:2006}. 
\end{example}

\subsubsection{Initialization}

A standard default initialization strategy for MCMC sampling is to
draw the initial parameter values from their priors. For Bayesian
finite mixture models this strategy is not advisable because such an
initial parameter vector $\bm{\vartheta}_K$ might have a very low
posterior probability leading to many burn-in iterations being
required to arrive at a parameter region with higher posterior
probability. In addition one usually also is at risk to be trapped in
a local optimum.

To obtain a better initial parameter vector, in particular in the
context of cluster analysis, the initial parameter values
$\bm{\eta}_K$ and $\bm{\Theta}_K$ may be obtained based on a partition
of the data set which is derived using some cluster analysis
method. E.g., $k$-means clustering can be applied to the data $\bm{y}$
with $k$-means clustering implicitly assuming equally sized clusters
and multivariate normal cluster distributions with spherical
variance-covariance matrices which are identical across clusters.

\begin{example}[Initialization]\label{ex:initialization}
  For the \code{diabetes} data set, the initial partition of the data
  into three groups is obtained by applying $k$-means clustering.  The
  initial values for the component means are set to the cluster means
  of the initial partition, whereas the initial variance-covariance
  matrices are set to $0.75 \bm{S}$ with $\bm{S}$ being a diagonal
  matrix containing the diagonal elements of $\textsf{COV}(\bm{y})$
  and the mixture weights set equal to $1/K$.
\end{example}

\subsubsection{Mixing and Convergence Diagnostics}

Using MCMC with data augmentation for fitting Bayesian mixture models
is known to be susceptible to poor mixing
\citep{Celeux+Hurn+Robert:2000}. In fact, the known multimodality of
the mixture posterior due to label invariance may be exploited to
assess convergence. Using label-invariant priors, the posterior takes
the same values regardless of how the labels are permuted. If a
mixture with $K$ components is fitted, this means that the posterior
has $K!$ identical modes induced by all possible permutations of the
labels. Thus a well mixing Markov chain should equally visit all these
modes and all posteriors of component-specific parameters should be
identical across $K$. To resolve the problem of partial label
switching and ensure that all these modes are equally visited,
\citet{Fruehwirth-Schnatter:2001} proposed the permutation sampler
which ensures balanced label-switching.

Visual inspection of the trace plots of the posterior draws of the
parameters as well as other quantities of interest such as the mixture
likelihood allow to assess convergence of the chain as well as the
presence of label switching in case non-label-invariant quantities are
inspected. Insights from these plots can be used to determine a
suitable number of burn-in iterations to be omitted and decide if
more iterations are required for suitable posterior inference.
If the visual inspection indicates that during MCMC sampling only one
mode of the mixture posterior is explored, there is in fact no need to
resolve label switching.  However, as label switching cannot be always
assessed visually, it is common practice to assume the presence of
label switching and resolve it in any case before performing
component-specific inference.

\begin{example}[Trace plots]
  
  The Gibbs sampling scheme outlined in
  Section~\ref{sec:model-estimation} and
  Example~\ref{ex:model-estimation} is used to draw $M=30000$ samples
  from the finite Gaussian mixture model specified in
  Examples~\ref{ex:data-model} and \ref{ex:hier-gener-model} to
  analyze the \code{diabetes} data set. All draws are retained without
  omitting any burn-in iterations in order to study the convergence
  behavior of the chain.  Figure~\ref{fig:trace} shows trace plots of
  the sampled component means of the first variable and the number of
  allocated observations to each of the components for the first 10000
  iterations. The figure indicates that the chain converged rather
  fast after about 1000 iterations. Also, it seems that no label
  switching occurred after convergence, at least for the plotted range
  of iterations. In the following the first 5000 iterations are
  omitted as burn-in and the remaining 25000 iterations used for
  posterior inference.
   \begin{figure}[h!]
     \centering
     \includegraphics[width=0.7\textwidth]{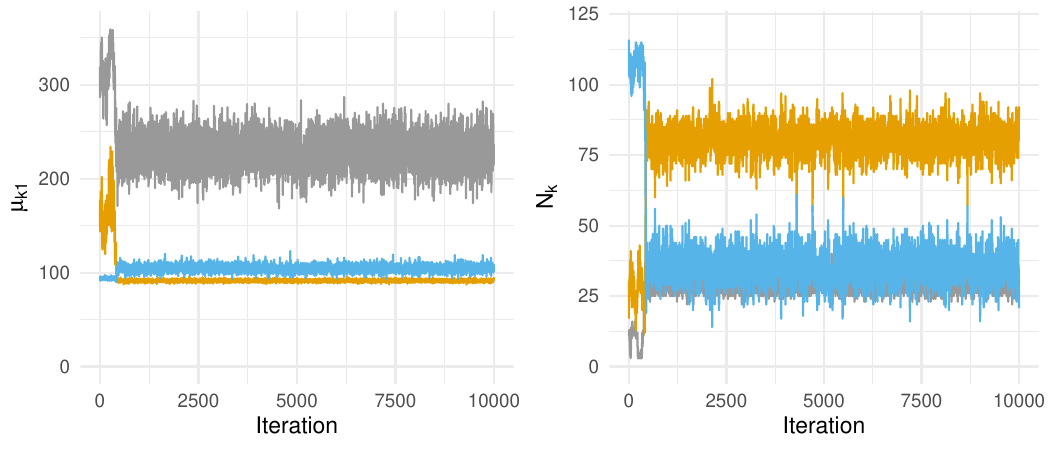}
     \caption{Trace plots of the $\mu_{k1}$ draws of the first
       variable (left) and the number of observations allocated to a
       component $N_k$ (right), in different colors for the three
       components.}
     \label{fig:trace}
  \end{figure}
\end{example}

\subsection{Post-Processing and Inference}\label{sec:post-proc-infer}

\subsubsection{Resolving label switching}

The approximations of the posterior distributions of parameters and
latent variables should be identical across the $K$ components in case
the chain has converged and equally visited all modes induced by the
$K!$ permutations of the labels. However, these approximations do not
help to characterize the component sizes and distributions or assign
observations to components. Hence, parameters which are
label-dependent need to be relabeled in order to focus on a single
mode of the posterior. Many different methods have been proposed to
obtain a unique labeling \citep[for an overview, see for
example][]{Papastamoulis:2016}.  For example, a unique labeling can be
achieved by imposing ordering constraints on the draws or by solving
an optimization problem to minimize the dissimilarity between the
classification matrices and some average of them.

Alternatively, a straightforward approach is to apply $k$-means
clustering to the point process representation of the MCMC draws to
identify the mixture model \citep{Fruehwirth-Schnatter:2006,
  Malsiner-Walli+Fruehwirth-Schnatter+Gruen:2016}.  We will refer to
this approach as the ppr approach in the following. The
component-specific parameters $\bm{\theta}_k$, $k=1,\ldots,K$, or some
(low-dimensional) functional, are clustered in the point process
representation into $K$ clusters using $k$-means clustering.  The
information which draws are from which component or from which MCMC
iteration is not included in the clustering procedure. Hence, even
though the draws are clustered into $K$ groups, not necessarily all
$K$ sampled component-specific parameter vectors of one MCMC sweep are
assigned to $K$ different clusters.  Only for those MCMC sweeps where
the $K$ sampled parameter vectors are clustered into exactly $K$
groups a unique labeling of the draws can be obtained. Thus, only
these MCMC draws are retained and the other MCMC draws are omitted
before relabeling and performing a component-specific analysis. The
proportion of MCMC sweeps which need to be omitted serves as indicator
how well the components can be distinguished. A high proportion of
omitted sweeps usually results if draws of different components are
very similar and therefore allocated to the same cluster by $k$-means.
This is in particular in a cluster analysis context an indication that
the fitted mixture model is overfitting and has too many components.

\begin{example}[Resolving label switching]\label{ex:resolv-label-switch}
  In order to obtain a unique labeling of the sampled draws, the
  sampled component means $\bm{\mu}_k$ are clustered into three groups
  using $k$-means. The bivariate scatter plots of the sampled
  component means $\bm{\mu}_k$ are shown in Figure~\ref{fig:ppr} in
  the first row. After applying $k$-means clustering the
  classification is shown in the second row. The sampled component
  means $\bm{\mu}_k$ are clearly allocated to three separate and
  distinct clusters. The non-permutation rate, i.e., the proportion of
  MCMC sweeps where the draws are clustered into less than three
  different clusters, is smaller than 0.01. Thus, for more than 99\%
  of the MCMC sweeps a unique labeling could be obtained and the
  remaining less than 1\% sweeps were excluded from further
  analysis. The obtained unique labeling is used to reorder not only
  the component mean draws $\bm{\mu}_k$, but also the draws of the
  other parameters, e.g., the mixture weights $\bm{\eta}_k$ and the
  assignment variables $\bm{S}$. After relabeling, component-specific
  inference is possible by, e.g., computing the posterior means. In
  Table~\ref{tab:comp} on the left the posterior means of the mixture
  weights and the component means are reported for the model identified
  using the ppr approach.
  \begin{figure}[h!]
    \centering
    \includegraphics[width=0.9\textwidth]{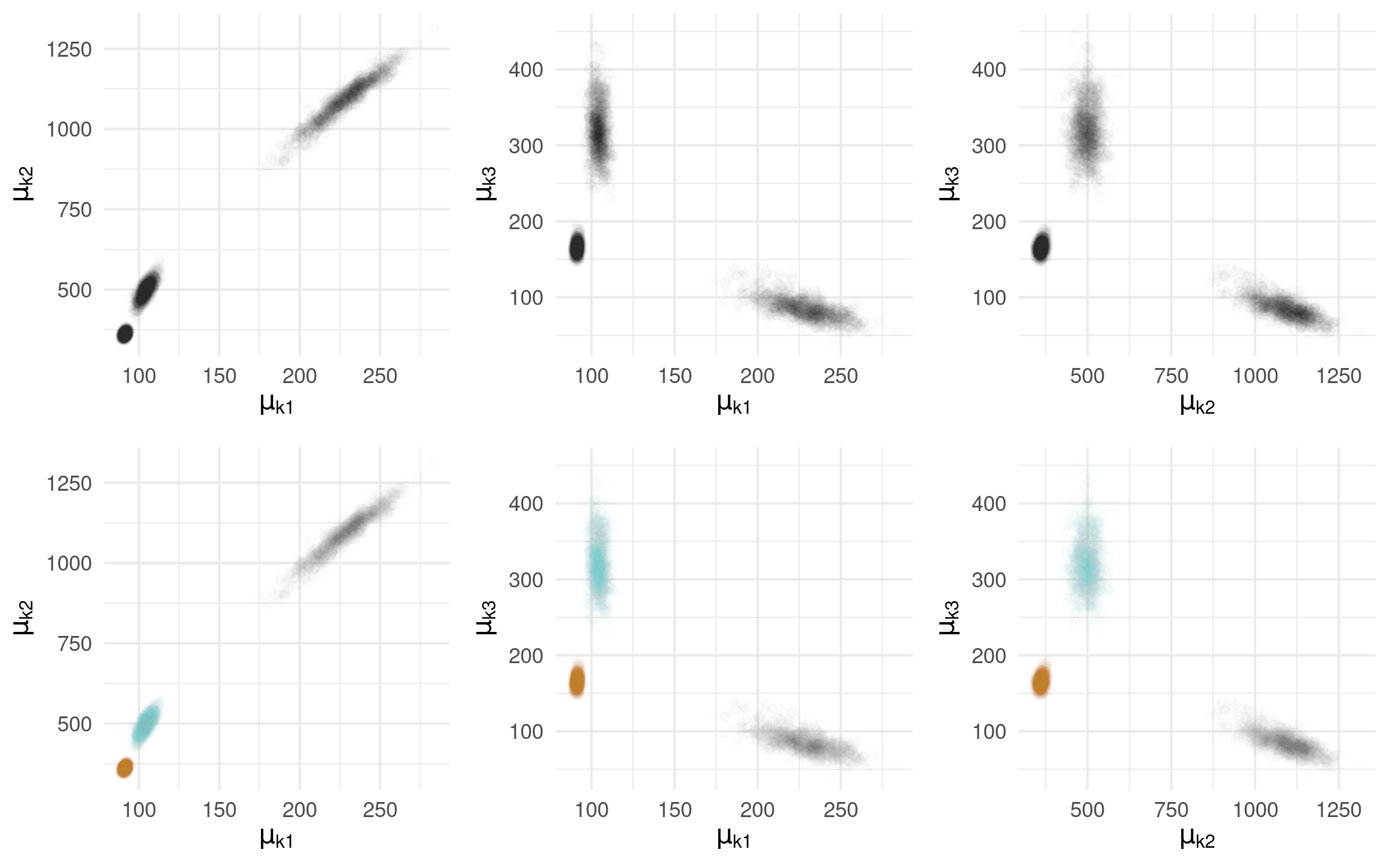}
    \caption{Bivariate scatter plots of the $\mu$ draws for all three
      components and for the three variables (on the top) and the
      retained $\mu$ draws after model identification with the ppr
      approach where the colors indicate the identified component
      labels (on the bottom). Only the first 1000 iterations are shown
      after discarding the first 5000 iterations as burn-in. }
    \label{fig:ppr}
  \end{figure}

\begin{table}
  \centering
  \caption{Component- or cluster-specific inference of the identified
    mixture models estimated with known $K = 3$ (left), using a sparse
    finite mixture (SFM) model with $K = 10$ and $\gamma = 0.01$
    (middle) and a dynamic mixture of finite mixtures (MFM) model with
    BNB$(1, 4, 3)$ prior on $K$ and $\gamma_K = 0.5 / K$ (right): Number
    of observations allocated to the components or clusters using the
    posterior mode of the assignment probabilities ($N_k$) and the
    posterior mean estimates of the component or cluster weights
    $\hat{\eta}_k$ and component or cluster means $\hat{\bm{\mu}}_k$,
    $k=1,2,3$.
    \label{tab:comp} }
  \begin{tabular}{rrrrcrrrcrrr}
    \toprule
    &\multicolumn{3}{c}{Known $K = 3$}&&\multicolumn{3}{c}{SFM with $K = 10$}&&\multicolumn{3}{c}{Dynamic MFM}\\
    \cmidrule{2-4}\cmidrule{6-8}\cmidrule{10-12}
    & 1 & 2 & 3 && 1 & 2 & 3 && 1 & 2 & 3\\ 
    \midrule
    $N_k$ & 28 & 33 & 84 &  & 28 & 33 & 84 &  & 28 & 33 & 84 \\ 
    $\hat{\eta}_k$ &    0.20 &    0.25 &    0.55 &  &    0.20 &    0.24 &    0.56 &  &    0.20 &    0.24 &    0.56 \\ 
    $\hat{\mu}_{k1}$ &  229.41 &  104.37 &   91.41 &  &  229.39 &  104.49 &   91.44 &  &  229.41 &  104.49 &   91.45 \\ 
    $\hat{\mu}_{k2}$  & 1098.04 &  496.87 &  361.43 &  & 1097.89 &  497.94 &  361.73 &  & 1097.97 &  497.78 &  361.89 \\ 
    $\hat{\mu}_{k3}$  &   82.66 &  319.27 &  165.19 &  &   82.72 &  321.17 &  165.47 &  &   82.71 &  321.89 &  165.44 \\ 
    \bottomrule
  \end{tabular}
\end{table} 
\end{example}

\subsubsection{Final partition}

After having resolved the label switching issue, the posterior of the
assignment vectors $\bm{S}$ can be used to assign the observations to
components and derive a partition of the data. For example, the
partition based on the maximum a-posteriori estimate is obtained if
each observation $\bm{y}_i$ is assigned to the component to which it
is allocated most often during MCMC sampling, i.e., based on the
posterior mode. Note that the partition derived in this way may
contain less than $K$ components.

Alternatively, if interest only lies in obtaining a final partition,
loss-based approaches can be used where the loss is label-invariant
and hence no unique labeling is required and thus resolving the label
switching can be avoided. This approach requires to solve an optimization
problem which consists of determining the partition which minimizes
the posterior loss \citep{Rastelli+Friel:2018}. Yet another
possibility is to estimate a similarity matrix between observations
based on their posterior co-allocation probabilities and then use
clustering methods to obtain a suitable partition based on the
similarity matrix. \citet{Molitor+Papathomas+Jerret:2010} apply
partitioning around medoids (PAM) to such a similarity matrix and
select a suitable number of clusters based on the silhouette width.

\begin{example}[Final partition]\label{ex:final-partition}
  The final partition of the data is obtained by relabeling the
  assignment vectors $\bm{S}$ according to the unique labeling
  obtained in the ppr approach and then assigning each observation to
  the component to which it was assigned most frequently during MCMC
  sampling. The resulting group sizes are reported in
  Table~\ref{tab:comp} on the left.  Table~\ref{tab:classification} on
  the left reports the confusion table of the estimated and true
  classification. The estimated cluster labels seem to be in line with
  the true class labels if both are sorted by group size and hence,
  the observations on the diagonal are assumed to be correctly
  classified. Based on the number of observations in the off-diagonal
  a misclassification rate (MCR) of 14\% is obtained (see
  Table~\ref{tab:classification} on the right). An additional measure
  to assess congruence between two partitions is the adjusted Rand
  index \citep[ARI;][]{Hubert+Arabie:1985}. The adjusted Rand index
  has the advantage that it is label invariant. The maximum value of
  one indicates perfect congruence between two partitions, the
  adjustment is used to obtain a value of zero if the number of
  co-allocations in the pair of partitions equals the number expected
  by chance. For the example data set the final partition estimated
  has an adjusted Rand index value of 0.65 (see
  Table~\ref{tab:classification} on the right). Determining the final
  partition based on the minimization of the posterior loss defined by
  the variation of information criterion \citep{Rastelli+Friel:2018}
  yields almost the same result with respect to the performance
  measures ARI and MCR. The final partition of the ppr approach is
  also visualized and compared to the true classification using
  bivariate scatter plots of the measurements in Figure~\ref{fig:classification}.
  
\begin{table}[t!]
  \centering
  \caption{Left: Confusion table of the true and the estimated
    classification obtained with the ppr approach.  Right: Partition
    congruence measures consisting of the adjusted Rand index (ARI)
    and the misclassification rate (MCR) for final partitions,
    obtained by the ppr method (ppr) or minimizing the posterior loss
    based on the variation of information method
    (VI).}\label{tab:classification}
  \begin{tabular}[c]{rrrr}
    \toprule
    &\multicolumn{3}{c}{Estimated}\\
    \cmidrule{2-4}
      True 	& 1 & 2 & 3 \\ 
    \midrule
    Overt &  27 &   6 &   0 \\ 
    Chemical &   1 &  24 &  11 \\ 
    Normal &   0 &   3 &  73 \\ 
    \bottomrule
  \end{tabular}
  \hspace*{1cm}
  \begin{tabular}[c]{lcll}
    \toprule
    Method & Groups & ARI & MCR \\
    \midrule
    ppr & 3 & 0.65 & 0.14 \\
    VI  & 3 & 0.64 & 0.15\\
    \bottomrule
  \end{tabular}
\end{table}
\begin{figure}[t!]
  \centering
  \includegraphics[width=0.9\textwidth]{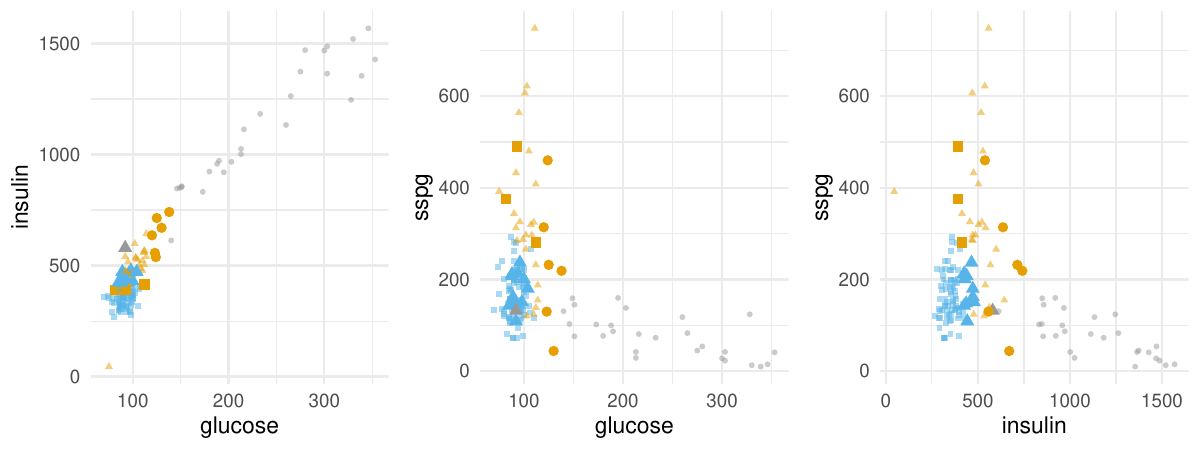}
  \caption{Bivariate scatter plots of the clinical measurements with
    the colors indicating the component membership estimated using the
    posterior modes of the component assignments after identification
    with the ppr approach and the shapes the true classes. The size of
    misclassified observations is
    increased.}\label{fig:classification}
\end{figure}
\end{example}

\section{Unknown Number of Components}\label{sec:unkown-numb-comp}

In many applications the number of groups is not known. In this case
data analysis does not only consist of fitting a mixture model with a
fixed number of components $K$, but usually also needs to provide
insights into how many data clusters are present in the data set.
\citet{Celeux+Fruehwirth-Schnatter+Robert:2019} point out that
determining the number of components to include in a finite mixture
model is not only an important question because inference about the
fitted model is highly sensitive to the $K$ selected. Determining $K$
is also most difficult due to the identifiability issues of the
mixture model and due to the fact that a suitable selection of $K$
might also depend on the modeling purpose of the mixture model.

An important distinction to make when estimating the number of
clusters in the data set at hand, is the conceptual difference between
the number of components $K$ in the finite mixture model and the
number of clusters in the data set, henceforth called $K_+$.
\citet{McCullagh+Yang:2008} discuss this distinction in the context of
species sampling where $K$ represents the number of species in the
population whereas $K_+$ corresponds to the number of species observed
in the available data set. These numbers need not to be the same as not
all species of the population might also be observed in the
sample. Correspondingly, not all available mixture components might be
used to generate the data at hand. Thus, it might be that $K_+<K$.

Maximum likelihood estimation of finite mixture models does not make
this distinction between $K$ and $K_+$, as the usual assumption is
that the number of clusters $K_+$ in the data exactly corresponds to
the number of mixture components $K$.  Problems with empty components
are usually only acknowledged if, for example, uncertainty assessment
based on bootstrapping is performed, see, e.g.,
\citet{OHagan+Murphy:Scrucca:2019} who propose to use a weighted
likelihood bootstrap to ensure that observations from all components
present in the original data set are also present in the bootstrap
samples.

In contrast, in Bayesian estimation of finite mixture models the
distinction between $K$ and $K_+$ is always present. This distinction
is of course made in Bayesian nonparametrics [\textbf{stat07850},
\textbf{stat07905}] where an infinite number of mixture components is
assumed while the data at hand only consists of a finite number of
clusters. However, also in finite mixture modeling the distinction
between $K$ and $K_+$ is natural. When assigning observations to the
components by drawing from the multinomial distribution with success
probabilities equal to the component weights, there is a-priori a
positive probability that not to all components observations are
assigned. This probability is the larger, the more unbalanced the
weights are. A-priori the balancedness of the weights is influenced by
their prior, i.e., the Dirichlet parameter $\gamma_K$.

Unfortunately, the handling of an overfitting mixture model where the
number of mixture components $K$ exceeds the number of clusters $K_+$
in the data set is not straightforward as identifiability problems
arise. Additional components can be added to any mixture model without
changing the induced mixture distribution: either a component weight
of zero is assigned to the new components or the new components are
obtained by duplicating an already existing component and splitting
the component weight between them.  \citet{Rousseau+Mengersen:2011}
provide theoretic insights into how the value $\gamma_K$ of the
Dirichlet prior on the mixture weights impacts on the asymptotic
behavior of the posterior of an overfitting finite mixture model.
\citet{Fruehwirth-Schnatter:2011} points further out how these
theoretic results can be used for practical data analysis. In fact, if
$K$ is not known and has to be estimated, $\gamma_K$ needs to be
specified in dependence of the model selection strategy pursued, as
will be explained in the following sections. The choice of $\gamma_K$
determines how large the gap between $K$ and $K_+$ is a-priori, i.e.,
the prior assumption on how many of the species in the population are
also observed in the data set.

In the following we discuss three strategies for Bayesian finite
mixture modeling when the number of components $K$ is unknown and one
aims at estimating the number of clusters $K_+$ in the data set. Note
that the problem of estimating the number of groups in a data set is
usually referred to as ``estimating $K$''. However, we explicitly
pursue a distinction between $K$ and $K_+$ and aim at estimating the
number of data clusters $K_+$, which may be different from $K$, the
number of mixture components or number of clusters in the population.

\subsection{Model Selection}\label{sec:model-selection}

If it is assumed that $K_+=K$, estimating $K_+$ can be viewed as a
model selection problem, where the most appropriate mixture model
within the class of mixture models with varying $K$ is selected.  The
posterior probability of each model is determined by calculating the
marginal likelihood in combination with a suitable prior on the
models. Such an approach has the advantage to not only allow
comparisons across mixture models with different number of components,
but also mixture models with different component distributions and
different prior specifications. Using equal prior model probabilities
the ratio between two posterior model probabilities corresponds to the
Bayes factor which also is often used as basis for model selection and
to compare different models.
  
If the model selection strategy aims at fitting finite mixture models
with different number of components $K$ and comparing their fit using,
for example, the posterior probabilities and the marginal likelihoods,
it is important to ensure that all components are filled and hence to
choose the Dirichlet parameter $\gamma_K$ large
\citep{Fruehwirth-Schnatter:2011}.  Determining the marginal
likelihood for finite mixture models is again complicated by the
multimodality due to label
switching. \citet{Fruehwirth-Schnatter:2019} suggests to estimate the
marginal likelihood with the full permutation or the double random
permutation bridge sampling estimator, however only for $K$ up to
seven.

\subsection{Sparse Finite Mixture Models}

A computational less intensive approach to estimate the number of
clusters in the data set is the ``sparse finite mixture'' (SFM)
approach.  Inspired by the theoretic insights provided by
\citet{Rousseau+Mengersen:2011},
\citet{Malsiner-Walli+Fruehwirth-Schnatter+Gruen:2016} propose the SFM
model where $K$ is selected large to induce on purpose an overfitting
finite mixture model. This model specification is combined with a
(very) small value of $\gamma_K$ such that for the superfluous
components the component weights a-priori and a-posteriori tend to
zero. This implies that a-priori a large gap between $K$ and $K_+$ is
imposed and the posterior of $K_+$ may flexibly adapt to the observed
clusters in the data.  The number of data clusters $K_+$ can be
estimated using the posterior mode of the number of filled components
induced by the sampled assignment vectors. In contrast to pursuing a
model selection strategy as explained in
Section~\ref{sec:model-selection}, fitting a sparse finite mixture
model does not require to fit finite mixture models with increasing
number of components $K$, but a single MCMC run of such an overfitting
mixture model is sufficient to estimate the number of data clusters
$K_+$.

If the posterior mode of the number of filled components is used to
select a suitable number of data clusters $K_+$, cluster-specific
posterior inference can be performed after omitting all MCMC sweeps
where more or less components were filled and retaining only MCMC
sweeps where exactly $K_+$ components were filled. Using the draws of
these MCMC sweeps first only the draws of component-specific
parameters associated with filled components are retained and those of
empty components omitted. These sweeps and draws are then used to
resolve label-switching and perform inference for the parameters
of the cluster-specific distributions.  The link between sparse finite
mixture models and Dirichlet process mixtures where $K$ is in fact
equal to infinity is empirically investigated in
\citet{Fruehwirth-Schnatter+Malsiner-Walli:2019}.

\subsection{Prior on the Number of Components}\label{sec:prionOnK}

In Bayesian analysis a natural approach to account for the unknown
number of components $K$ is to introduce a prior on $K$ and include
the determination of the posterior of $K$ as well as the posterior of
$K_+$ in the Bayesian inference.
The hierarchical generative model given in (\ref{eq:BFM}) then has the
additional layer at the top:
\begin{align*}
  K &\sim p(K),
\end{align*}
where $p(K)$ is the prior on the number of components. Following
\cite{Miller+Harrison:2018} we call the resulting model a mixture of
finite mixtures (MFM)
model. \citet{Fruehwirth-Schnatter+Malsiner-Walli+Gruen:2020} review
previous suggestions for the prior on $K$ and propose to use the
beta-negative-binomial (BNB) distribution for $K - 1$ to ensure that
the support is equal to the positive integer numbers. The BNB
distribution is a three-parameter distribution which allows to not
only calibrate the prior mean and variance, but also take into account
the probability for the homogeneity model $K = 1$ and the tail
behavior, i.e., the mass assigned to large values of $K$. The BNB
distribution has the Poisson, geometric and negative-binomial
distribution as special cases.

The mixture of finite mixtures model has been investigated by
\citet{Richardson+Green:1997} with inference based on the reversible
jump MCMC for mixtures of univariate Gaussians.  Specific extensions
of this reversible jump MCMC algorithm to also cover other component
distributions were considered in
\citet{Zhang+Chan+Wu:2004} and
\citet{Papastamoulis+Iliopoulos:2009}. As an alternative to reversible
jump MCMC \citet{Stephens:2000} proposes a sampling scheme based on
creating a Markov birth-death process.  Further
\citet{Miller+Harrison:2018} suggest to re-use methods from Bayesian
nonparametrics and apply the Jain-Neal sampler developed for Dirichlet
process mixtures \citep{Jain+Neal:2004,Jain+Neal:2007}. In this
sampling scheme the number of components $K$ is not included but each
observation in turn is re-assigned to an already filled component or a
new component during MCMC sampling, thus effectively sampling
partitions of the data and hence $K_+$. To fit the mixture of finite
mixtures model, \citet{Fruehwirth-Schnatter+Malsiner-Walli+Gruen:2020}
propose the ``telescoping sampler'' which extends the usual MCMC
scheme with data augmentation by including also a step where $K$ is
explicitly sampled conditional on the current partition of the data
and another step to obtain parameters for the empty components:
\begin{enumerate}
\item[3.] Sample $K$ conditional on current cluster sizes $(N_1,\ldots,N_{K_+})$ and $\gamma_K$ from
\begin{eqnarray*}
  p(K|(N_1,\ldots,N_{K_+}),\gamma_K) &\propto
	\frac{K!}{(K-K_+)!}
	\frac{\Gamma(K\gamma_K)}{\Gamma(K \gamma_K +N)}	 
	\prod_{k=1}^{K_+} \frac{\Gamma(N_k+ \gamma_K)}{\Gamma(1+ \gamma_K)} p(K).
\end{eqnarray*}
\item[4.] Add $K-K_+$ empty components with component-specific parameters
sampled from the priors.
\end{enumerate}
Step~2 is modified to draw the mixture weights only after the empty
components were added and the hyper-parameters are drawn conditional
on the component-specific parameters of the filled components,
disregarding the empty components.

This sampling scheme is very generic allowing for arbitrary component
distributions by just re-using sampling schemes already available for
finite mixtures with a fixed number of components $K$. In contrast to
\citet{Richardson+Green:1997} and \citet{Miller+Harrison:2018} the
telescoping sampler is also applicable with arbitrary values selected
for $\gamma_K$. \citet{Fruehwirth-Schnatter+Malsiner-Walli+Gruen:2020}
consider in particular the ``dynamic'' specification of a mixture of
finite mixtures model where $\gamma_K=\alpha/K$, i.e., the Dirichlet
parameter decreases with increasing number of components $K$ implying
a larger gap between $K$ and $K_+$ for larger values of $K$.

\begin{example}[Fitting a SFM and MFM model]\label{ex:prior-numb-comp}
  
  A sparse finite mixture model with $K=10$ mixture components and
  $\gamma=0.01$ is fitted to the data set. The sampler is run for
  $M=30000$ iterations. The trace plot of the number of components $K$
  (in gray) and filled components $K_+$ (in black) is shown in
  Figure~\ref{fig:SFM+MFM}a) on the left. After starting with an
  initial partition of 10 filled components, rather fast most of them
  become empty and the sampler switches mostly between three and four
  filled components. On the right, the relative frequencies of the
  different $K_+$ values sampled after omitting a burn-in of 5000
  iterations is plotted. A clear mode at $K_+=3$ is discernible. Thus
  three clusters are estimated for the data set and all MCMC sweeps
  where the number of filled components is not equal to three are
  discarded. After discarding also the draws sampled from empty
  components the ppr approach is used for the remaining draws to
  identify the three clusters. In Table~\ref{tab:comp} (middle) the
  posterior means of the cluster means and cluster weights of the
  identified model are reported as well as the group sizes of the
  final partition. The results correspond to those when fitting a
  mixture model with exactly three components and the same final
  partition is obtained.

  \begin{figure}[t!]
    \centering
    \includegraphics[width=0.49\textwidth]{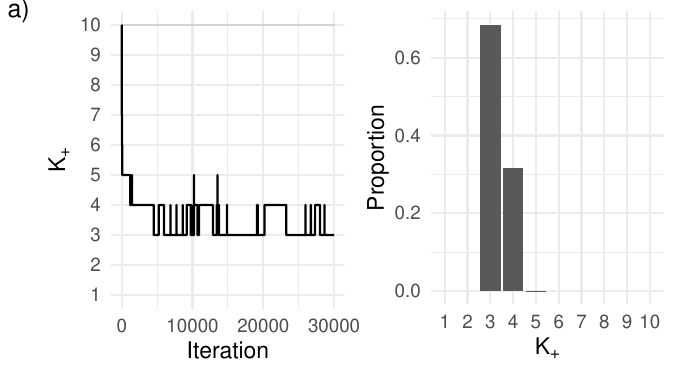}
    \hfill
    \includegraphics[width=0.49\textwidth]{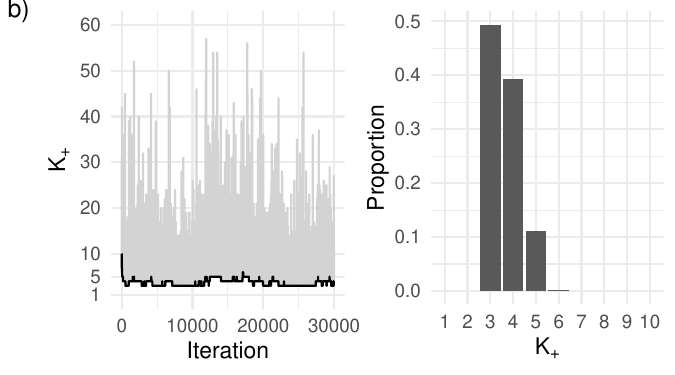}
    \caption{a) Sparse finite mixture with $K=10$, $\gamma=0.01$. b)
      Dynamic MFM with $K\sim \text{BNB}(1,4,3)$, $\alpha=0.5$. For each
      modeling approach, on the left, trace plot of the number of
      components (gray) and the number of filled components (black)
      and on the right, bar plot of the relative frequencies of the
      number of filled components.}\label{fig:SFM+MFM}
  \end{figure}
 
  Finally, a dynamic MFM model with the prior
  $K-1\sim \text{BNB}(1,4,3)$ and $\gamma_K=0.5/K$ is fitted to the
  data using the telescoping sampler. The trace plots of the sampled
  $K$ (in gray) and induced $K_+$ (in black) are shown in
  Figure~\ref{fig:SFM+MFM}b) on the left. The sampled $K$ values
  assume large values, even greater than 20. The sampled $K_+$ values
  again switch between three and four filled components.  The bar plot
  in Figure~\ref{fig:SFM+MFM}b) on the right visualizes the relative
  frequencies of $K_+$ values sampled after omitting 5000 burn-in
  iterations.  Again the mode of the posterior of $K_+$ is at $K_+=3$,
  although not so clear as when fitting a SFM model. After
  identification of the model for $K_+=3$, the posterior means of the
  cluster-specific parameters and the group sizes of the final
  partition are reported in Table~\ref{tab:comp} (right). Again the
  results obtained correspond to those for $K$ fixed to three and the
  exact same final partition is obtained.  
\end{example}

\section{Concluding Remarks}\label{sec:concluding-remarks}

Recent research has provided insights into the specification and
estimation of Bayesian finite mixture models including the impact of
different priors and improvements in estimation and inference, in
particular regarding computational issues. In the future we would
expect the routine application of Bayesian finite mixture models in
model-based clustering in different areas of application as well as
using data with ``unusual'' structure. This will induce the need to
include various component-specific distributions or models. We have
focused on mixtures of distributions in this paper. However, clearly
also a regression setting can be included where the components differ
with respect to the regression coefficients or mixture-of-expert
models where the component weights also depend on additional
covariates. Recent advances might also help to use Bayesian finite
mixture models for semi-parametric density
approximation.

\section*{Acknowledgments}

The authors gratefully acknowledge support from the \textit{Austrian
  Science Fund (FWF)}: P28740; and through the \textit{WU Projects}
grant scheme: IA-27001574.



\bibliography{bayesian-mixtures}

\end{document}